\documentstyle[psfig]{lamuphys}
\makeatletter
\let\chapter\hid@chapter
\makeatother
\newcommand{\kms}{{km s$^{-1}$}\/}
\newcommand{\as}{{$''\!\!$.}\/}
\newcommand{\HST}{{\sl HST}\/}
\begin{document}

\index{spectrum!nuclear}
\index{variability!nuclear}

\pagenumbering{arabic}
\title{The Nuclear Spectrum of M87}

\author{Z.I.\,Tsvetanov\inst{1}, 
        G.F.\,Hartig\inst{2}, 
        H.C.\,Ford\inst{1,2},
        G.A.\,Kriss\inst{1}, 
        M.A.\,Dopita\inst{3}, \\
        L.L.\,Dressel\inst{4}, and
        R.J.\,Harms\inst{4}}

\institute{Johns Hopkins University, Baltimore, MD 21218, USA
\and
Space Telescope Science Institute, Baltimore, MD 21218, USA
\and
Mount Stromlo and Siding Spring Observatories, ACT 2611, Australia
\and
RJH Scientific, 5904 Richmond Highway, Alexandria, VA 22303, USA}

\maketitle

\begin{abstract}

The nuclear spectrum of M87 covering the Ly$\alpha$-H$\alpha$
wavelength range was obtained with the \HST\ Faint Object Spectrograph
(FOS) trough a 0\as21 aperture. Contrary to some previous claims, a
single power law ($F_{\nu} \sim \nu^{-\alpha}$) can not reproduce the
observed continuum shape and at least a broken power law is required
for a good fit ($\alpha = 1.75$ and $1.41$ shortward and longward of
the break at $\sim$4500 \AA). We detect a set of broad (FWHM $\sim$
400 \kms) absorption lines arising in the gas associated with
M87. These are only lines from neutral and very low ionization species
blueshifted by $\sim$150 \kms\ relative to the M87 systemic velocity,
indicating a net gas outflow and turbulence.  The excitation sensitive
emission line ratios suggest that shocks may be the dominant energy
supplier.
 
The nuclear source in M87 is significantly variable.  From the FOS
target acquisition data, we have established that the flux from the
optical nucleus varies by a factor $\sim$2 on time scales of $\sim$2.5
months and by as much as 25\% over 3 weeks, and remains unchanged
($\leq$ 2.5\%) on time scales of $\sim$ 1 day.  These timescales
limit the physical size of the emitting region to a few hundred
gravitational radii. The variability, combined with other observed
spectral properties, strongly suggest that M87 is intrinsically of BL
Lac type but is viewed at an angle too large to reveal the classical
BL Lac properties.
 
\end{abstract}

%\vspace{-3mm}
\section{Introduction}
%\vspace{-1mm}

M87 with its famous synchrotron jet is, perhaps, one of the most
studied extragalactic objects in modern astronomical research. In this
respect the spectrum of the nucleus is of particular importance for
mapping the properties of the central source and a better
understanding of its physics.  Nevertheless, there is basically no
good spectrum of the nucleus in the UV-optical-IR region.  This is at
least partially due to the difficulty of isolating from the ground the
central point source from the bright host galaxy light. The best
nuclear spectroscopy has concentrated on obtaining high resolution,
high S/N ratio data for measuring the central black hole (BH) mass.

The situation changed dramatically with the installation of corrective
optics (COSTAR) on \HST\ in December 1993.  Several critical
observations have been made since then. These include the discovery of
a nuclear gaseous disk (Ford et al.\ 1994) and vastly improved
measurements of the BH mass (see Ford et al.\ and Macchetto et al.\ in
these Proceedings) among others.

Here we present the first spectrum of the nucleus of M87 obtained
solely with the Faint Object Spectrograph on \HST\ though a small
0\as21 aperture. At \HST\ resolution the flux in this aperture is
entirely dominated by the nuclear point source, and the contribution
from the host galaxy can be neglected. The Ly$\alpha$--H$\alpha$
coverage allows us to study the continuum shape and map the absorption
and emission line spectrum significantly better than any previous
attempt. In accordance with several recent works we assume a distance
to M87 of 15 Mpc and a systemic velocity of 1280 \kms.

\vspace{-3mm}
\section{The nuclear spectrum}
\vspace{-2mm}
\index{spectrum!nuclear}

\subsection{Continuum shape}
\vspace{-1mm}
\index{spectrum!nuclear!continuum}

The observed nuclear spectrum is dominated by a smooth featureless
continuum with broad emission lines and a number of absorption lines
imprinted on it (Fig.~1). We do not see any starlight features above
the noise and estimate that the stellar contribution does not exceed
the 10\%--15\% level at optical wavelengths.

\begin{figure}[ht]
%\vspace{5.0cm}
\vspace{-18mm}
\centerline{\psfig{figure=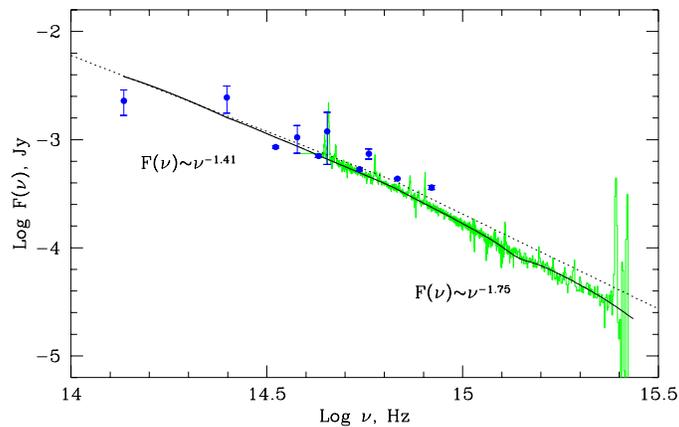,width=9.0cm}}
\vspace{-20mm}
\caption{The observed spectrum (light gray line) is plotted binned to
the FOS resolution (1 diode) to reduce the noise. Our best fit -- a
broken power law with $E(B-V) = 0.039$ -- is shown by a solid line. The
two intrinsic (i.e., unreddened) power laws are shown as dashed lines
connecting at the break of $\sim 4500$ \AA. The filled circles are
broad-band measurements from BSH91, Ze93 and St97. Our fit is
extrapolated into the IR to show its position relative to the
ground-based measurements. \label{fig:cont} }
\vspace{-3mm}
\end{figure}

Several earlier studies (e.g., Thomson et al.\ 1995) have strongly
argued that the nuclear spectrum is synchrotron emission. Some authors
have suggested that either optical-IR-radio (Zeilinger, Peletier \&
Stiavelli 1993) or optical-UV-X-ray (Biretta, Stern \& Harris 1991)
portions could be represented by a single power law. All these
studies, however, relied on broad-band ground-based optical and IR
photometry of the nuclear source.

In fitting the continuum shape we have adopted the strategy of finding
the simplest possible model producing an acceptable fit to the
data. The foreground galactic reddening is estimated at $E(B-V)$ =
0.017 (Jacoby, Ciardullo \& Ford 1990). In addition, the M87 nuclear
gaseous structure contains dust (see Sparks, Ford \& Kinney 1993 and
$\S$2.2 below), and thus there will be some additional extinction if
our line of sight to the nucleus passes through it.

The basic result from our fitting is that the observed continuum shape
is inconsistent with a single power law regardless of the amount of
extinction and the reddening law. There is additional curvature in the
spectrum that can be accounted for by introducing a break in the power
law (see Fig.~\ref{fig:cont}). Of course, a smooth, curved model
similar to the synchrotron self-Compton solutions employed by
Stiavelli, Peletier \& Carollo (1997) will also be a good
representation of our data. In this respect, it is important to note
that extrapolation of our best-fit broken power law misses both the
radio and X-ray observations by a very significant margin.

\vspace{-2mm}
\subsection{Absorption and emission lines}
\vspace{-1mm}
\index{spectrum!nuclear!absorption lines}

\begin{figure}
%\vspace{6.0cm}
\vspace{-3.0mm}
\centerline{\psfig{figure=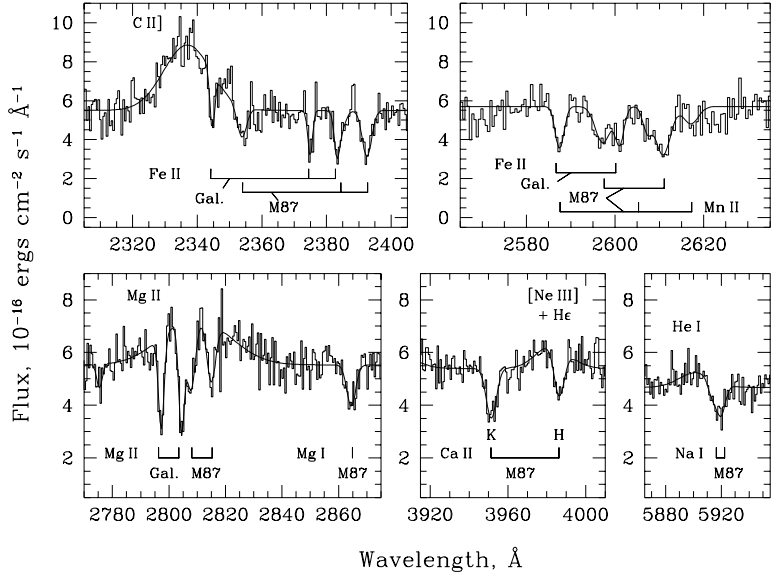,width=11.5cm}}
\vspace{-37mm}
\caption{Selected wavelength regions from the observed spectrum
   (histogram) and fitted Gaussian models (solid line). Absorption 
   lines are labeled below the spectrum, and emission lines -- above 
   it. Note that M87 absorption lines are broader than Galactic ones 
   (which remain unresolved at the FOS resolution). \label{fig:a-lines} }
\vspace{-3mm}
\end{figure}

We detect two clearly separated absorption line systems -- one arising
in the interstellar gas of our own galaxy and another one in that of
M87. Figure \ref{fig:a-lines} shows the selected wavelength ranges
with fitted Gaussian models overplotted onto the observed spectrum.
We fitted all lines in each group simultaneously, imposing a minimum
set of restrictions.

The averaged observed heliocentric velocity of the absorption lines
arising in M87 is 1202 $\pm$ 22 \kms, or 1134 $\pm$ 22 \kms\ if
corrected for the suspected wavelength zero point offset inferred from
the comparison of positions of the Galactic and H~{\sc i} absorption
features. This means that the gas responsible for the absorption in
M87 is blueshifted by $\sim$150 \kms\ relative to the systemic
velocity of 1280 \kms. 

In M87 we see absorption only from neutral and very mildly ionized gas
-- all ions have ionization potential $\chi < 10$ eV. In addition all
M87 absorption lines are broad with FWHM $\sim$ 400 \kms. At the same
time the Galactic absorption lines are unresolved by the FOS (FWHM
$\sim$ 220 \kms). We suggest that the M87 absorption line properties
can be understood in a simple model where our LOS to the nucleus
passes through an outflow from the inclined ($\sim 30^{\circ}$) and
turbulent nuclear disk.

\index{spectrum!nuclear!emission lines}
A significant number of emission lines are clearly visible in the
entire wavelength region covered by our spectrum. These include both
permitted and forbidden transitions of ions with a large range of
ionization potentials -- H~{\sc i}, He~{\sc i}, He~{\sc ii}, O~{\sc
i}, O~{\sc ii}, O~{\sc iii}, C~{\sc ii}, C~{\sc iii}, C~{\sc iv},
Ne~{\sc iii}, S~{\sc ii}, Mg~{\sc ii}. With no exceptions all lines
are very broad -- the best isolated and high S/N ratio lines such as
[O~{\sc iii}] $\lambda$5007 and H$\beta$ have FWHM $\sim$ 2000 \kms,
and other line widths are consistent with this value. This is an
expected result given the size of the FOS aperture (0\as21 = 15 pc)
and the broadening due to the Keplerian rotation in the gaseous disk
surrounding the central supermassive black hole (see Ford et al.\ and
Macchetto et al.\ in these Proceedings).

Another important characteristic of the M87 nuclear emission line
spectrum is its LINER type. In fact, the spectrum of the nucleus is
quite similar to the spectrum of the gaseous disk discussed in detail
by Dopita et al.\ (these Proceedings). Here we note only that
diagnostic diagrams involving critical UV emission line ratios clearly
indicate shock heating is the dominant excitation mechanism. For more
details see the contribution by Dopita et al. 

\vspace{-3mm}
\section{Variability}
\vspace{-2mm}
\index{variability!nuclear}

\begin{figure}[ht]
%\vspace{4.0cm}
\vspace{-6.5mm}
\centerline{\psfig{figure=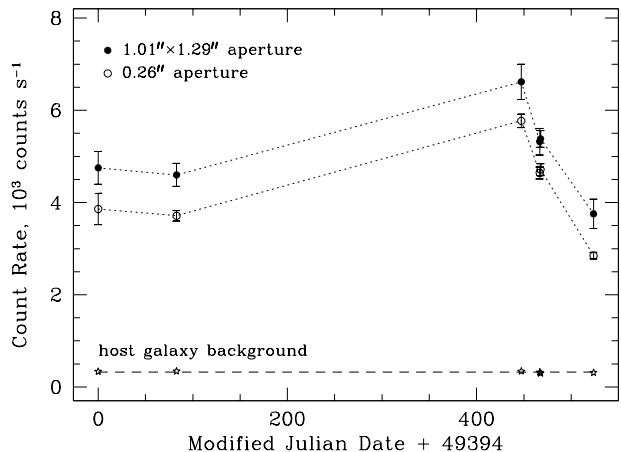,width=9.5cm}}
\vspace{-30mm}
\caption{Time dependence of the flux from the unresolved nucleus 
    measured from the binary search data (filled circles) and from 
    the peakup series (open circles) and the underlying galaxy (stars). 
    The 1$\sigma$ error bars are about 5\% and 2.5\% for the BS and 
    PU data, respectively, and are smaller than the symbols for the 
    galactic component. The extraction aperture is 1\farcs01 $\times$ 
    1\farcs29 for the BS and 0\farcs26 for the PU data. The constant 
    BS/PU flux ratio is characteristic of a point source for the 
    aperture difference (Keyes et al.\ 1995). \label{fig:var} }
\vspace{-3mm}
\end{figure}

A full presentation of the detected variability of the M87 optical/UV
nucleus is given in Tsvetanov et al.\ (1998). Measurements of the
nuclear flux and the host galaxy at 6 epochs in the 1994 February --
1995 August time period are shown in Fig.~\ref{fig:var}. Here we
summarize the basic results.

Analysis of the target acquisition data yields the following results:
(1) The nucleus remains unresolved at the \HST\ resolution.  The
characteristic size of the FOS/COSTAR point-spread function imposes an
upper limit of $\sim$60 mas (= 5 pc) in diameter to the nuclear
source size. (2) The flux from the nucleus changes by a factor $\sim$2
over 2.5 months, $\sim$25\% over 3 weeks, and remains the same to
within the errors ($\le$2.5\%) during 1 day. (3) The continuum
spectrum becomes bluer as it brightens while emission lines remain
unchanged.

Some important conclusions can be drawn from the above results. (i)
The one-month characteristic variability time scale and finite light
travel time combined with the independently estimated gravitational
radius of the BH in M87 put an upper limit to the size of the emitting
region of $l \le 200 R_{\rm g}$. (ii) The changes by a factor $\sim$2
indicate that at least half of the nuclear flux is variable. Yet, the
total energy output in the variable component is only a small fraction
of the Eddington limit, as is the total power released by the BH.
(iii) The detected variability, combined with the observed continuum
shape, relativistic boosting, and significant superluminal motions
(Biretta et al.\ these Proceedings) imply strongly that M87 is
intrinsically of BL Lac type but is viewed from an angle too large
to make its BL Lac properties dominate.

%
% ---- Bibliography ----
%

\end{document}